\documentclass[final]{aipproc}

\layoutstyle{6x9}

\def\lsi{\raise0.3ex\hbox{$<$\kern-0.75em\raise-1.1ex\hbox{$\sim$}}}
\def\gsi{\raise0.3ex\hbox{$>$\kern-0.75em\raise-1.1ex\hbox{$\sim$}}}

\newcommand{\gsim}{\mathop{\gsi}}
\newcommand{\R}{{\kern+.25em\sf{R}\kern-.78em\sf{I} \kern+.78em\kern-.25em}}
\newcommand{\RR}{{\kern+.25em\sf{R}\kern-.6em\sf{I} \kern+.6em\kern-.25em}}

\begin{document}

\title{Chiral Fermions on the Lattice}


\author{Wolfgang Bietenholz}{
  address={Instituto de Ciencias Nucleares, Universidad Nacional
  Au\'{o}noma de M\'{e}xico, \\ A.P. 70-543, C.P. 04510 Distrito Federal,
  Mexico}
}

\begin{abstract}
In the last century the non-perturbative regularization
of chiral fermions was a long-standing problem.
We review how this problem was finally overcome by the
formulation of a modified but exact form of chiral symmetry
on the lattice. This also provides a sound definition of the
topological charge of lattice gauge configurations. 
We illustrate a variety of applications to
QCD in the p-, the $\epsilon$- and the $\delta$-regime, where
simulation results can now be related to Random Matrix Theory 
and Chiral Perturbation Theory. The latter contains Low
Energy Constants as free parameters, and we comment
on their evaluation from first principles of QCD.
\end{abstract}

\maketitle

\vspace{-7mm}
\section{Chiral symmetry}

\paragraph{Chiral Perturbation Theory}

Fermion fields can be decomposed into a left- and a right-handed
component by means of the chiral projectors,
$\Psi_{L,R} = \frac{1}{2} ( 1 \pm \gamma_{5}) \Psi \,$, 
$\bar \Psi_{L,R} = \bar \Psi \frac{1}{2} ( 1 \mp \gamma_{5}) \,$.
In massless (bilinear) theories these two spinor components decouple.
In particular the QCD Lagrangian at zero quark masses takes the
structure
\begin{equation}
{\cal L}_{\rm QCD} = \bar \Psi_{L} \, D \, \Psi_{L} + 
\bar \Psi_{R} \, D \, \Psi_{R} + {\cal L}_{\rm gauge} \ ,
\end{equation}
where $D$ is the Dirac operator, and the quark fields 
$\bar \Psi, \ \Psi$ capture the $N_{f}$ flavors and 3 colors 
involved. This ${\cal L}_{\rm QCD}$ is invariant under global 
$U(N_{f})$ transformations of the quark fields, which can be 
performed independently in the left- and right-handed sector.
The two complex phases represent baryon number
conservation and an axial symmetry, which breaks under
quantization (axial anomaly). One assumes the remaining
{\em chiral flavor symmetry} to break spontaneously, reducing
the symmetry to a unbroken group of simultaneous transformations
in both sectors,
\begin{equation} \label{XSB}
SU(N_{f})_{L} \otimes SU(N_{f})_{R} \to SU(N_{f})_{L+R} \ .
\end{equation}
{\ Chiral Perturbation Theory} ($\chi$PT) deals with a
field in the corresponding coset space, $U \in SU(N_{f})$,
which dominates the low energy behavior.
As we add small quark masses $m_{q}$ to ${\cal L}_{\rm QCD}$
(which is allowed, since QCD is a vector theory),
$U$ represents $N_{f}^{2} -1$ light quasi-Nambu-Goldstone bosons, 
which are identified with the lightest mesons involved. 
For simplicity we consider only two (degenerate) flavors, 
$u$ and $d$, so that the field $U$ represents the pion 
triplet.\footnote{In this case, the symmetry breaking pattern
(\ref{XSB}) is locally isomorphic to $O(4) \to O(3)$. 
This property is very specific:
among all conceivable types of chiral symmetry breaking only 
very few can be expressed by orthogonal Lie groups \cite{XSBnonU}.} 
$\chi$PT now uses an effective Lagrangian of the form \cite{XPT}
\begin{equation} \label{Lxpt}
{\cal L}_{\rm eff}[U] = \frac{F_{\pi}^{2}}{4} {\rm Tr} 
[ \partial_{\mu} U^{\dagger}\partial_{\mu} U ] -
\frac{\Sigma m_{q}}{2} {\rm Tr} [U + U^{\dagger}] + \dots
\end{equation}
where the dots represent terms with more derivatives
and/or higher powers of the explicit symmetry breaking parameter
$m_{q}$. All terms, which are compatible with the symmetries, are put
into an energetic hierarchy; eq.\ (\ref{Lxpt}) displays the
leading terms. Each term comes with a coefficient, which is denoted
as a {\em Low Energy Constant} (LEC), such as the {\em pion decay 
constant} $F_{\pi}$ (which can be measured experimentally) and the
{\em chiral condensate} $\Sigma$ (the order parameter of chiral
symmetry breaking). 

As we have seen, $\chi$PT does have a
direct link to QCD, hence it describes low energy hadron physics 
in a way manifestly related to the fundamental theory (in contrast to
many other effective approaches). However, the LECs are free
parameters in $\chi$PT; in this sense the low energy description is
incomplete. {\em If} we manage to determine LECs directly from QCD, 
we obtain a more complete low energy theory.
This is obviously a non-perturbative task, and therefore a challenge
for {\em lattice QCD}: its simulation in Euclidean space
is the only method to tackle QCD (and other quantum fields theories)
systematically beyond perturbation theory. 

\paragraph{Lattice fermions}

The lattice discretization of the gluon fields is conceptually 
unproblematic: the gauge action can be expressed in terms of
small Wilson loops in a gauge invariant way. It has been a 
longstanding issue, however, to formulate lattice fermions
such that they keep track of (approximate) chiral symmetry.
For one flavor, the standard chirality condition is given 
by the anti-commutator $\{ D , \gamma_{5} \} = 0$.
The ``na\"{\i}ve'' discretization of the Dirac operator yields for
a free, massless fermion in momentum space the form
$D_{\rm n}(p) = {\rm i} \gamma_{\mu} \sin p_{\mu}$ (in lattice 
units, {\it i.e.}\ if we set the lattice spacing $a=1$).
It is chirally symmetric, but it gives rise to $2^{d}-1$ 
artificial poles of the propagator (inside the first Brillouin 
zone), in addition to the physical one at $p=0$ (in $d$ dimensional
Euclidean space). 
The Nielsen-Ninomiya Theorem states essentially that chirality
and locality inevitably entail fermion doublers, 
which would distort the result in lattice studies \cite{NoGo}. 
(Here ``locality'' means that the coupling between $\bar \Psi_x$ and
$\Psi_y$ falls off at least exponentially in $|x-y|$; this assures 
a safe continuum limit).

K.\ Wilson subtracted a discrete Laplacian $\Delta$ 
to construct the Wilson Dirac operator \cite{Wilfer}, 
$D_{\rm W} = D_{\rm n} - \frac{1}{2} \Delta$.
It is still local, and it sends the doubler masses to the cutoff 
scale, as desired. However, the additional term breaks 
chiral symmetry explicitly. Under gauge interaction it 
leads to (highly undesired) additive mass 
renormalization. Thus the chiral limit can only be approximated
by a tedious fine-tuning of a negative bare quark mass.

Conceptual progress was achieved at the end of the last century
by deviating from the continuum form of chiral symmetry in a 
specifically harmless way: instead of inserting a local term
for $\{ D , \gamma_{5} \}$ (as Wilson did), one now does so for
$\{ D^{-1} , \gamma_{5} \}$ --- this does not shift the poles in
the propagator --- {\it e.g.}\ by setting
\begin{equation} \label{GWR}
\{ D^{-1} , \gamma_{5} \} = \gamma_5 \quad \Rightarrow \quad 
\{ D , \gamma_{5} \} = D \gamma_{5} D \ .
\end{equation}
This is now known as the (simplest form of the) {\em Ginsparg-Wilson
Relation} (GWR). Lattice Dirac operators with this structure are 
generated by block spinor Renormalization Group Transformations as the
blocking factor diverges ({\em perfect fermion action})
\cite{GiWi,QuaGlu,Has97}, or by {\em Domain Wall Fermions} 
\cite{DWF} (with the chiral modes attached to
two domain walls, which are pulled apart in an extra ``dimension'').
By integrating out this extra ``dimension'' one obtains the
{\em overlap fermion} \cite{Neu1}. The latter can easily 
be re-derived from the GWR (\ref{GWR}): assume $D$ to be 
$\gamma_{5}$-Hermitian, $D^{\dagger} = \gamma_5 D \gamma_{5}$
(which holds {\it e.g.}\ for $D_{\rm W}$), so that the GWR
can be written as $D + D^{\dagger} = D^{\dagger} D$. This 
corresponds to the condition that $A := D -1$ be unitary.
For $A_{\rm W} = D_{\rm W} -1$ this is not the case, but we can 
enforce it by the transformation
\vspace*{-4mm}
\begin{equation}  \label{overlap}
A_{\rm ov} = A_{\rm W} / \sqrt{A_{\rm W}^{\dagger} A_{\rm W} } \ , 
\qquad D_{\rm ov} = A_{\rm ov} + 1 \ ,
\end{equation}
which yields the ``overlap Dirac operator'' $D_{\rm ov}$ as a 
solution to the GWR \cite{Neu2}. Its spectrum is located on a circle
in the complex plane, $|\lambda -1| = 1$,
which reveals the absence of additive mass renormalization. 
(This also holds for the generalization to $A = D - \rho$,
$|\lambda -\rho | = \rho$, 
$\rho \gsim 1$, which has practical advantages in the interacting case.)
The overlap operator is manifestly local as long as the gauge
background is sufficiently smooth \cite{HJL}. This property gets
lost on very coarse lattices ($a \gsim 0.17 ~ {\rm fm}$), 
but if we simulate in the safe regime, 
a continuum extrapolation can be taken.

Zero modes of a Ginsparg-Wilson Dirac operator
are exact and they have a definite chirality.
Hence the topological charge $\nu$ of a gauge configuration
can be defined \cite{HLN} by adapting the Atiyah-Singer Index Theorem
from the continuum. 
Thus $\nu$ is defined as the difference between the number
of zero modes with positive and negative chirality. (In
actual Monte Carlo generated configurations only zero modes with
one chirality occur.)
In contrast to the formulation using $D_{\rm W}$, 
operator mixing (on the regularized level) 
is under control \cite{Hasmix}. This is very helpful for numerical 
measurements, for instance if one performs 
a fully non-perturbative Operator Product Expansion \cite{OPE}. 

As a reason for these fantastic properties, M.\ L\"{u}scher
pointed out that the Lagrangian is actually invariant under 
a lattice modified, chiral transformation with the infinitesimal 
form \cite{ML98}
\vspace*{-3mm}
\begin{equation}
\bar \Psi \to \bar \Psi \Big( 1 + \varepsilon \Big[ 1 - \frac{1}{2}
D \Big] \gamma_{5} \Big) \ , \quad 
\Psi \to \Big( 1 + \varepsilon \gamma_{5} 
\Big[ 1 - \frac{1}{2} D \Big] \Big) \Psi \quad ({\rm to} 
~ O(\varepsilon )) \ .
\end{equation}
In the continuum limit it turns into the standard
chiral transformation,
$\bar \Psi \to \bar \Psi (1 + \varepsilon \gamma_{5})$,
$\Psi \to (1 + \varepsilon \gamma_{5}) \Psi$.
However, the fermionic Lagrangian $\bar \Psi D \Psi$ is invariant to
$O(\varepsilon )$ even on the lattice, {\em if} the GWR holds.
On the other hand, the functional measure ${\cal D} \bar \Psi \,
{\cal D} \Psi$ is {\em not} invariant,
which is exactly what it takes to reproduce the axial
anomaly correctly \cite{ML98,DA}.

For applications also further properties --- beyond chirality --- matter.
This motivates the substitution of the Wilson kernel $D_{\rm W}$ in the 
overlap formula (\ref{overlap}) by an extended ``hypercube fermion'',
which is constructed from Renormalization Group Transformations
and truncations \cite{EPJC}; this yields improved locality and
scaling, as well as approximate rotation symmetry \cite{WBIH,WBStani}.
In particular locality now persists on coarser lattices, which
is profitable in studies of QCD at finite temperature \cite{Biel}.

\section{Applications to topology and p- , $\epsilon$- and 
$\delta$-regime}

We mentioned that the topological charge
is well-defined when we deal with chiral lattice
fermions (although all lattice gauge configurations can be
continuously deformed into one another).
This enables a sound numerical measurement of the
{\em topological susceptibility} 
$\chi_{t} = \langle \nu^{2} \rangle / V$,
where $\nu$ is the (aforementioned) topological charge, and $V$ is
the space-time volume. A high statistics study was presented
in Refs.\ \cite{chitop}, which is very well compatible with our
results \cite{WBStani}, see Figure 1 (left). The continuum 
extrapolation amounts to $\chi_{t} = (191(5) ~ {\rm MeV})^{4}$.
This supports the {\em Witten-Veneziano scenario} 
\cite{WiVe}, which explains the heavy mass of the $\eta'$-meson 
in part as a topological effect. This conjecture involves indeed the
{\em quenched} value of $\chi_{t}$.
Latest direct measurements of $m_{\eta}$ and $m_{\eta '}$ were
performed with 2+1 flavors of dynamical Domain Wall quarks \cite{NC}.\\

$\chi$PT has been formulated in different regimes depending on the
volume, which affects the counting rules for the energy hierarchy.

\paragraph{p-regime}

The standard setting, where finite size affects are small,
is denoted as the {\em p-regime} \cite{preg}: $L \gg 1/m_{\pi}$ 
($L$ is the 4d box length, and the inverse pion mass is the 
correlation length).
Figure 1 (right) shows a measurement of $F_{\pi}$ that
we performed (quenched) on a $12^{3} \times 24$ lattice, where 
the gauge coupling was chosen such that the lattice spacing
corresponds to $a \simeq 0.123 ~ {\rm fm}$. The pion mass
(and other hadron masses) can be measured from the exponential
decay of correlation functions.
Over a broad range, the measured value of $F_{\pi}$ is 
clearly too high, but just at our lowest pion mass,
$m_{\pi} = 279(32) ~ {\rm MeV}$, the right trend sets in,
{\it i.e.}\ a decrease
towards the value in Nature ($F_{\pi} = 92.4 ~ {\rm MeV}$).
This calls for clarification with yet lighter pion masses,
still closer to its phenomenological value of $135 ~ {\rm MeV}$.
However, at our lowest data point we already have $L m_{\pi} \approx 2$,
hence at even smaller $m_{\pi}$ we are certainly outside
the p-regime. Recovering it takes a much larger volume,
and therefore much more computational effort.

\paragraph{$\epsilon$-regime}

As an alternative, we can simulate QCD in the {\em $\epsilon$-regime}
\cite{epsreg1}, where  $L m_{\pi} < 1$. This setting is unphysical --- 
experimentally we can't squeeze pions into such a tiny box.
However, the finite size effects can be computed by $\chi$PT,
and they are parameterized by {\em LECs as they occur in infinite
volume}. Hence we can extract physical results nevertheless,
avoiding the apparent quest for a huge volume \cite{GHLW}.
In the $\epsilon$-regime the topological sectors play an
essential r\^{o}le. 

\begin{figure}
  \includegraphics[angle=270,width=.5\linewidth]{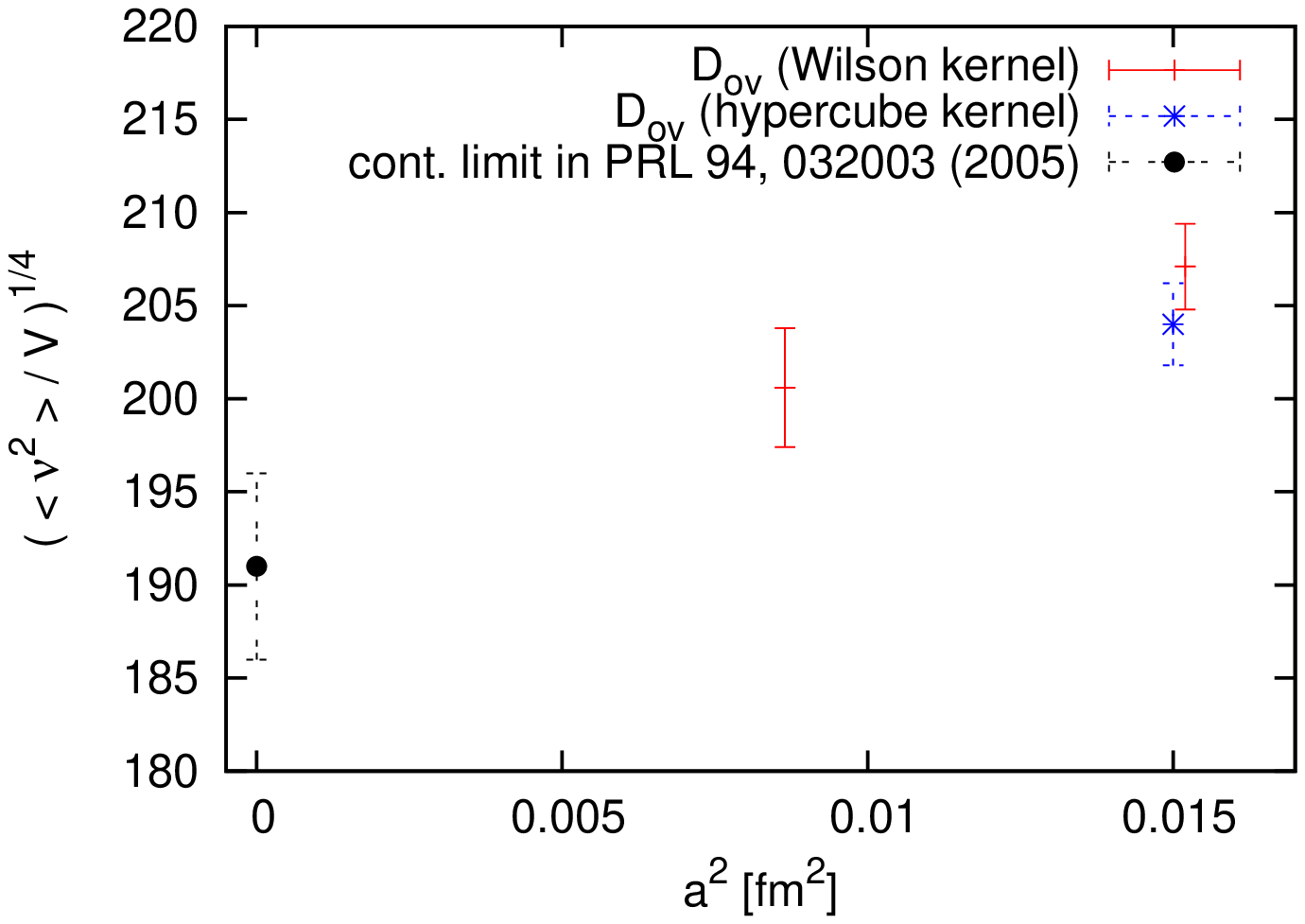}
  \includegraphics[angle=270,width=.5\linewidth]{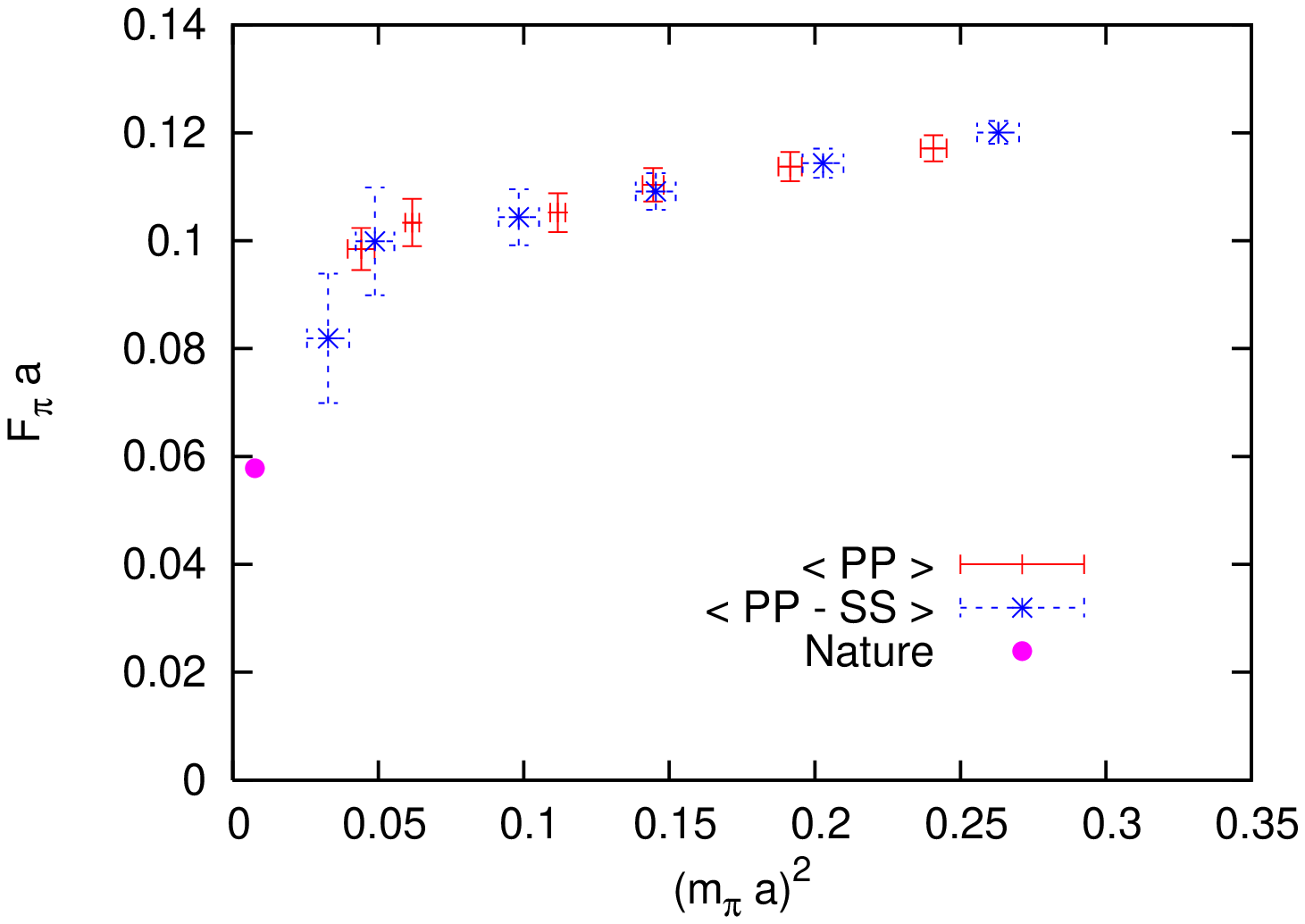}
\vspace*{-4mm} \\
  \caption{{\em Left:} The topological susceptibility of quenched
QCD, measured with the index of chiral quarks. 
Our results \cite{WBStani}
are consistent with the continuum limit of Refs.\ \cite{chitop},
which supports the Witten-Veneziano scenario for $m_{\eta'}.$
{\em Right:} $F_{\pi}$ measured by two methods (straight 
pseudoscalar density correlator, and subtraction of scalar
density correlator) at different pion masses in the p-regime
(at $a \simeq 0.123 ~ {\rm fm}$). 
For our lightest pion, $m_{\pi} = 279(32) ~ {\rm MeV}$, a trend 
towards the value in Nature sets in \cite{WBStani}.
\vspace*{-5mm}}
\end{figure}

Chiral Random Matrix Theory (RMT) provides a prediction for the 
density of the lowest non-zero Dirac eigenvalues $\lambda_{i}$ 
($i=1,2,3 \dots$) in the $\epsilon$-regime \cite{lowEV}.
More precisely it predicts the densities of the dimensionless
variables $z_{i} = \lambda_{i} \Sigma V$. 
If our data match the predicted shape, we can tune 
$\Sigma$ for optimal agreement, and in this way evaluate $\Sigma$.
Figure 2 (left) shows the cumulative densities for $z_{1}$ as
predicted by RMT in the sectors of topological charge
$|\nu | = 0, \ 1$ and $2$ (curves). Our data points, obtained in 
$V \simeq (1.23 ~ {\rm fm})^{4}$, are in excellent agreement,
if we insert $\Sigma \simeq (253 ~{\rm MeV})^{3}$ \cite{RMT}. 
Also this result was obtained in the quenched approximation
(which neglects sea quark contributions),
but our result showed for the first time that this method to 
measure $\Sigma$ is in fact applicable with chiral fermions in
various topological sectors --- see also Refs.\ \cite{RMT2,WBStani}.
Recent studies with dynamical quarks \cite{JLQCD} obtain with the same 
method a very similar value, $\Sigma = (251(7) ~{\rm MeV})^{3}$ 
(renormalized in the $\overline{MS}$ scheme at $2 ~{\rm GeV}$) .

However, simulations of dynamical overlap fermions are not
only computationally very expensive (the inverse square
root in eq.\ (\ref{overlap}) has to be computed by polynomials
up to degree $O(100)$ in order to attain chirality
close to machine precision), but they also face
conceptual problems: the standard algorithm for dynamical
quarks (``Hybrid Monte Carlo'') changes the
topological sector only very rarely, so that direct
measurements of full observables are difficult.
Measurements can be performed in fixed topological
sectors. A method to derive from them an
approximate result for the physical value (properly
summed over all sectors) has been suggested in Ref.\ \cite{BCNW} and
tested successfully in the 2-flavor Schwinger model \cite{WBIH08}.
The tremendous efforts to simulate QCD with dynamical overlap
quarks are reviewed in Ref.\ \cite{Hide}.
In particular $\chi_{t}$ is hard to measure in this case;
for an indirect method we refer to Refs.\ \cite{Japchit}.

\begin{figure}
  \includegraphics[angle=270,width=.5\linewidth]{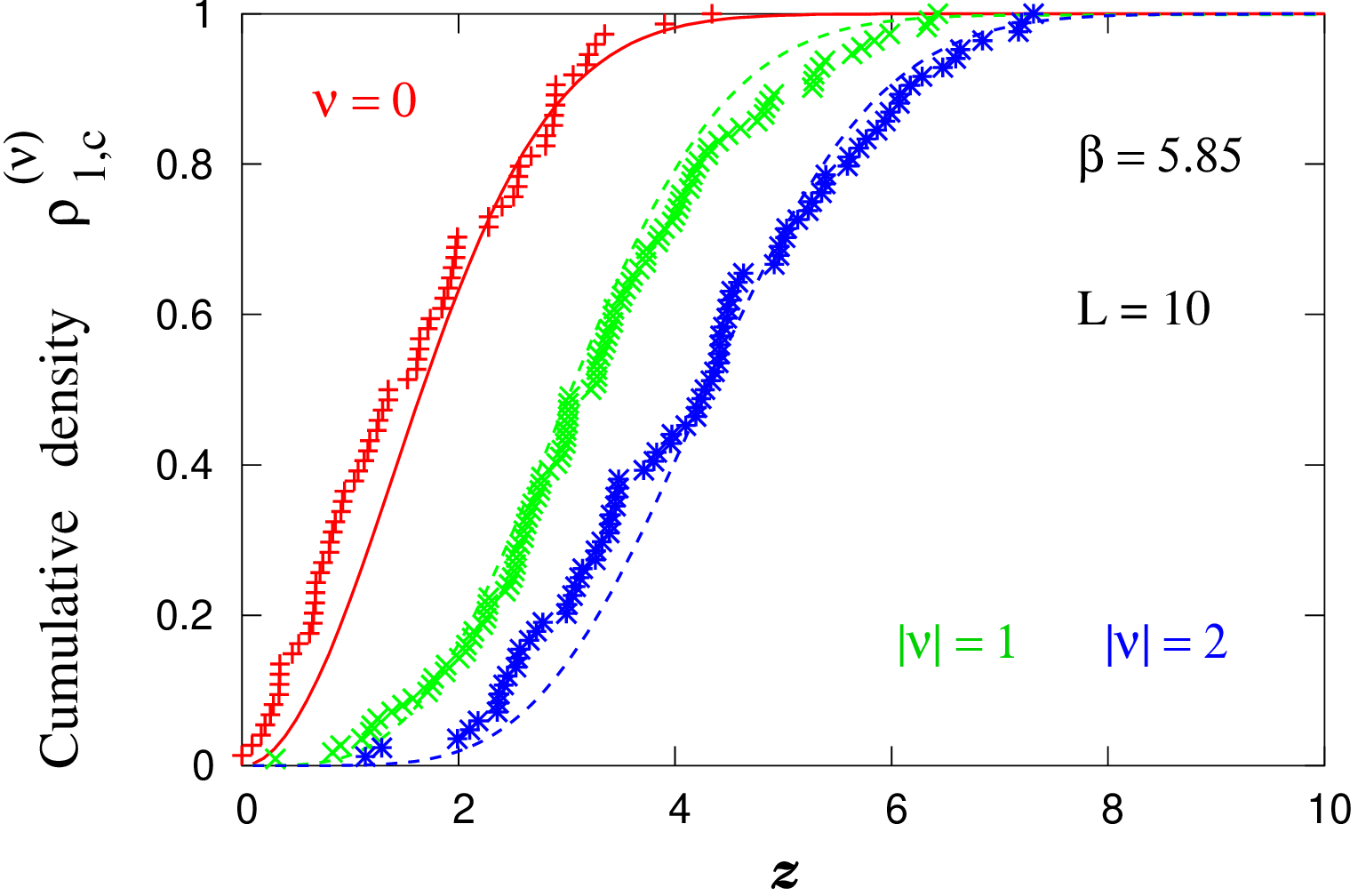}
  \includegraphics[angle=270,width=.5\linewidth]{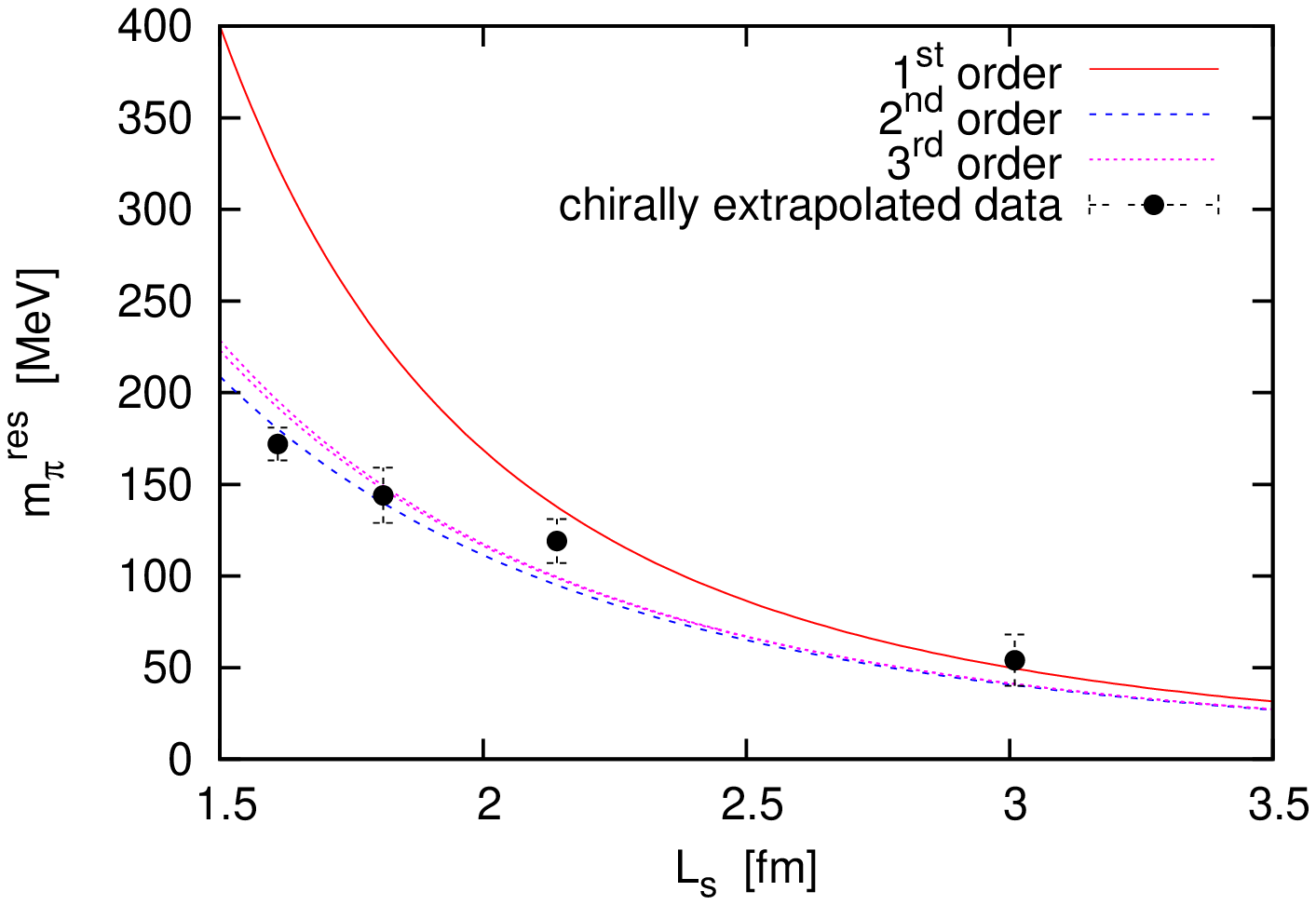}
  \caption{{\em Left:} The cumulative density of the dimensionless
variable $z_{1} = \lambda_{1} \Sigma V$, where $\lambda_{1}$ is the
leading non-zero Dirac eigenvalue. 
We compare our data for $D_{\rm ov}$ eigenvalues in the
$\epsilon$-regime (mapped stereographically onto $\R_{+}$)
to RMT predictions (lines) in the sectors with topological charge
$|\nu | = 0, \ 1$ and $2$. We obtain very good agreement if we insert
$\Sigma = (253 ~{\rm MeV})^{3}$ \cite{RMT}. {\em Right:}  The residual 
pion mass in the  $\delta$-regime as a function of the spatial box
size $L_s \,$. The chiral extrapolation of our measured pion masses
(with dynamical Wilson quarks) \cite{QCDSFdelta} follows closely the 
theoretical prediction of $\chi$PT \cite{Has3}.
\vspace*{-7mm}}
\end{figure}

We add that also $F_{\pi}$ can be evaluated in the 
$\epsilon$-regime, in particular by matching measured
correlators \cite{AApap,WBStani} to $\chi$PT 
predictions, or by considering only their zero-mode 
contributions \cite{zeromodes,WBStani}.

\paragraph{$\delta$-regime}

Let us finally mention yet a third regime where $\chi$PT
has been worked out, namely the $\delta$-regime \cite{deltareg}. 
Here the Euclidean time extent is long, but the 3d spatial volume,
say $L_{s}^{3}$, is small ($L_{s} < 1/m_{\pi}$).
This prevents spontaneous symmetry breaking,
hence even at vanishing quark mass --- where we can refer
to the current quark mass measured through the PCAC relation ---
the pion mass remains finite. The formula for the {\em residual 
pion mass} $m_{\pi}^{\rm res}(L_{s})$ in 
the chiral limit has been computed recently to next-to-next-to-leading
(NNL) order \cite{Has3}.\footnote{This is very different from the 
p-regime, where finite size effects are suppressed exponentially 
\cite{Lusch}.} 

We performed p-regime measurements of $m_{\pi}$ in spatial
volumes in the range $L_{s} \simeq (1.6 \dots 3.0) ~ {\rm fm}$,
and extrapolated the pion masses to the chiral limit \cite{QCDSFdelta}. 
The results are in remarkably good agreement with 
the formula for $m_{\pi}^{\rm res}(L_{s})$. The latter involves
$F_{\pi}$ again, so in principle this is yet another way to 
measure a physical LEC in a unphysical regime (although in our case
we already used $F_{\pi}$ for the extrapolation).

Moreover the NNL order also involves {\em sub-leading LECs} 
of $\chi$PT (coefficients to terms symbolized with dots in 
eq.\ (\ref{Lxpt})). Hence from precision results for
$m_{\pi}^{\rm res}(L_{s})$ 
in the $\delta$-regime one could determine even sub-leading LECs.
That is useful in particular for the LEC denoted as 
$\bar l_{3}$ \cite{QCDSFdelta}, the value of which is
quite uncertain \cite{CGL}.

\vspace{-2mm}
\section{Conclusions}

Chiral fermions can be regularized on the lattice such that
they obey a lattice modified version of chiral symmetry.
Therefore they are now well-defined and tractable 
non-perturbatively, at least in vector theories.
Thus the existence of light quarks --- with masses 
far below $\Lambda_{\rm QCD}$ --- is not that mysterious anymore
(for reviews, see {\it e.g.}\ Refs.\ \cite{revGWR}).
This formulation also provides a sound definition of a topological 
charge, which enables a neat measurement of the topological 
susceptibility in quenched QCD. The results support the 
Witten-Veneziano conjecture about the $\eta'$-mass.


We sketched applications of chiral lattice fermions in the p-regime, 
where finite size effects are small. Here we can measure 
for instance the light hadron spectrum as well as the PCAC quark mass, 
and $F_{\pi}$ --- one of the leading 
LECs in the $\chi$PT Lagrangian. The LEC determination from the underlying
theory (QCD) improves the status of $\chi$PT as a description of the
low energy hadronic world in a way linked to first principles.

In the $\epsilon$-regime we discussed the measurement
of $\Sigma$ --- the other leading LEC --- by relating the
microscopic Dirac spectrum to Random Matrix Theory. Also here, 
and in the $\delta$-regime, $F_{\pi}$ can be measured, {\it i.e.}\
it is possible to obtain physical results even from
unphysically small volumes. Moreover the measurement of the
residual pion mass in the  $\delta$-regime even has the
potential to determine sub-leading LECs. 

\vspace{2mm}
{\bf Acknowledgments :}
I thank my collaborators in the works summarized here, and the
organizers of the pleasant workshop in Mazatl\'{a}n.

\bibliographystyle{aipproc}   

\end{document}